# Edge Computing: A Comprehensive Survey of Current Initiatives and a Roadmap for a Sustainable Edge Computing Development

Andrea Hamm[1,2], Alexander Willner[2,3], Ina Schieferdecker[1,2]

[1] Weizenbaum Institute for the Networked Society, Berlin, Germany
[2] Technical University Berlin, Berlin, Germany
[3] Fraunhofer Institute FOKUS, Berlin, Germany
```
andrea.hamm@tu-berlin.de,
alexander.willner@fokus.fraunhofer.de,
ina.schieferdecker@web.de
```

**Abstract.** Edge Computing is a new distributed Cloud Computing paradigm in which computing and storage capabilities are pushed to the topological edge of a network. However, various standards and implementations are promoted by different initiatives. Lead by a reference architecture model for Edge Computing, current initiatives are analyzed by explorative content analysis. Providing two main contributions to the field, we present, first, how current initiatives are characterized, and second, a roadmap for sustainable Edge Computing relating three dimensions of sustainable development to four cross-concerns of Edge Computing. Findings show that most initiatives are internationally organized software development projects; important branches are currently telecom and industrial sectors; most addressed is the network virtualization layer. The roadmap reveals numerous chances and risks of Edge Computing related to sustainable development; such as the use of renewable energies, biases, new business models, increase and decrease of energy consumption, responsiveness, monitoring and traceability.

# 1 Introduction

Edge Computing allows to carry out the computing task on the spot; for example, during the opening ceremony of the 2018 PyeongChang Olympic Winter Games, Edge Computing approaches were used in the Olympic flag light show flown by drones [1]. Edge Computing describes a decentralized way of computing and storage at the topological edge of the network, i.e. in physical proximity to the data sourcing device. In contrast to the Cloud Computing paradigm, where the collected data is fully transmitted to a central server before being analyzed and used, Edge Computing allows to do these tasks on devices or close to it. Decentralized computing itself is not a new paradigm and has been prevalent before the centralized cloud computing paradigm. Now, it becomes more important as volumes of data production, traffic, analysis, and storage increase drastically, while bandwidth as well as security are limiting factors. Edge Computing further allows for deterministic behavior, which would not be realizable by communicating over the best-effort Internet [2]. The Edge Computing paradigm can further be understood as a distributed Cloud Computing paradigm and as such as a supplement to current Cloud Computing approaches.

Previous works emphasized an important role of Edge Computing for several sectors such as automated manufacturing [3], healthcare systems [4], telecommunication [5], or autonomous traffic systems [6]. Through the continuous advancements of ubiquitous computing, wireless sensor networks and machine-to-machine communication (M2M), connected devices are growing in number. Heterogeneous physical devices are enabled to transfer signals over the internet and become uniquely identifiable [7]. In particular, the Industrial Internet of Things (IIoT), i.e. the application of IoT approaches in industrial fields of application, urges for privacy-aware and low latency data processing [8]. For example, concepts of the *Smart Factory* or *Industry 4.0* often demand deterministic low-latency communication as large parts of the production processes are highly synchronized with each other [9]. The development of the IIoT has a large economic impact as new solutions are demanded, while conventional architectural patterns and infrastructures are challenged [8]. Other examples include the telecommunication and automotive industries that are pushing for low latency and reliable communications for innovations around connected cars [10] and other use cases, under the heading of multi-access edge computing (MEC). Edge Computing has been also identified as an enabling technology for numerous use cases in the development of the Internet of Things and the $5^{th}$ Generation Network (5G) [11] and is therefore an important piece to study.

A recent ACM article urges for a computing within limits that "should be paying serious attention to futures in which we encounter planetary limits." [12, p.93]. Within the Information Systems field, sustainability is an emerging topic for several years [e.g. 13, 14]. Today the issue becomes even more pressing with regard to the climate change debate. In this way, the German Advisory Council on Global Change (WBGU) summarizes among other institutions that today's greatest challenge of technological developments is to meet the natural resource scarcity of our planet.[15].

Following this principle, we propose a sustainability roadmap for Edge Computing. It should serve the developer's and manager's communities in the Edge Computing field to include sustainability dimensions, such as the usage of distributed renewable energy sources for edge nodes, as concerns that are as important as the technical concerns ensuring the functionality and efficiency of Edge Computing.

This study combines several research fields, i.e. computing development, social science methods, and sustainability studies. Due to its transdisciplinary character this article cannot elaborate on all research standards of the three disciplines. Nevertheless, it represents an important step of knowledge combination and shows multiple links for further research. We provide two main contributions to Edge Computing research: first, a comprehensive survey of current Edge Computing initiatives worldwide to overview current working areas and understand the state of the art; and second, a roadmap that relates sustainable development dimensions to current technical concerns of the Edge Computing development process.

A limitation of the study is that the analyzed sample is not exhaustive and therefore of limited representativity. The sample is probably biased towards the German context as one of the search terms "Industry 4.0" originates from a German policy.

The remainder of the paper is structured as follows: In Section 2, we start by related work in the area of Edge Computing and elaborate in particular on a reference architecture model for Edge Computing that has been suggested by the Edge Computing Consortium Europe (ECCE)[1]. In Section 3, we explain the research design to survey current Edge Computing initiatives by explorative content analysis. In Section 4, the results are presented and explained. In Section 5, we discuss the results regarding potential issues and gaps of the paradigm development and propose a roadmap for a more sustainable development of Edge Computing. The paper ends with a conclusion in Section 6.

## 2    Related Work

### 2.1    Cloud Computing and Sustainable Development

During the past decades Cloud Computing has tremendously increased the amount of transferred data and communications. From 2008 to 2017, the worldwide IP traffic per month has grown from 10,127 to 122,000 petabytes [16, 17], which is a boost by factor 12. Due to the fast-growing Internet of Things, it is predicted that the IP traffic will even more increase up to 396,000 petabyte in 2022 [17]. Terms like "the cloud" depicting a fluffy formation with no weight or mass, are obfuscating the fact that most of such digital solutions consume massive digital infrastructures with additional electricity demands, which are often times provided by old-fashioned and carbon-emitting energy supplies [18]. The synchronization of systems, where the end device might be located on the other side of the world than the data center, has caused a high increase of energy consumption at the sites of the central servers [19]. In an expected case scenario, the electricity consumed by communication technology will increase from about 2,000 TWh/year in 2010 to more than 8,000 TWh/year in 2030. This scenario includes energy consumption of data centers, consumer devices, WiFi and wired fixed access, and wireless networks access. Approaches to carbon-neutralize this energy consumption by investing in green energies or by building renewable energy plants close to a data center are not productive for sustainable development since the overall energy demand remains increased through these data services. If only the additional demand will be carbon-neutralized, the general energy demand, e.g. on a country basis, will not decrease.

---

[1] https://ecconsortium.eu

**2.2	Edge Computing Paradigm**

The Edge Computing paradigm is a multi-dimensional problem space. Next to the abundant technical challenges and approaches, there are numerous economic and political interests involved such as the transformation of information and communication technologies (ICT) markets or the strengthening of a country's digital economy. For these reasons, the field of Edge Computing is highly complex and suffers from inconsistent terms, standards, norms, and understandings. During the past years, several initiatives tackling different aspects of Edge Computing appeared. Among them collaborations between companies on an international level, community-based software projects or software initiatives started by companies. Roman et al. [20] presented an overview of the partly simultaneously existing Edge Computing paradigms encompassing e.g. fog computing, mobile edge computing, mobile cloud computing, and other approaches. With the early emergence of the decentralized computing paradigm, several standards related to today's understanding of Edge Computing have been registered or are currently under consideration. Some examples include:

- IEC 61131: Standard for programmable controllers
- IEC 61499: Standard for distributed programmable controllers
- IEC 62541: OPC Unified Architecture (OPC UA) as a communication middleware for industrial automation
- IEEE 1934: Standard for Adoption of OpenFog Reference Architecture for Fog Computing
- IEEE 802.1: Time-Sensitive Networking (TSN) Task Group related standards
- ETSI MEC: Multi-access Edge Computing standard
- DIN SPEC 92222: Reference model for the industrial cloud

**2.3	Industry 4.0 as an Application Domain**

Edge Computing applications are increasingly demanded in the field of Industry 4.0. Lasi et al. [21] describe Industry 4.0 as a vision for the future industry, containing two paradigm changes, first, on a technological level, and second, for the business culture. The technological changes would concern the emergence of fully digitized manufacturing environments, an extreme increase of actuator and sensor data and connected technical components, as well as the miniaturization of factories due to acceleration of computing power in today's devices. The changes in business culture encompass the acceleration and flexibility of the innovation cycle, an individualized production according to demands, an increased decentralization and reduction of hierarchies.

Additionally, Industry 4.0 would demand a greater resource efficiency to meet resource scarcity and requirements of sustainable development. Several studies have taken into account the potentials of Industry 4.0 for sustainable development [22, 23]. Pointing out that the complex business processes behind Industry 4.0 would not develop to an ecological relief by themselves, it must be consciously striven for such and actively shaped in the Industry 4.0 management [22]. In this study, we use the same approach for the Edge Computing paradigm as it is going to be an important aspect of Industry 4.0. By directly including sustainability concerns into the further edge development process, we aim to contribute likewise to a more sustainable future industry.

## 2.4    Reference Architecture Model for Edge Computing (RAMEC)

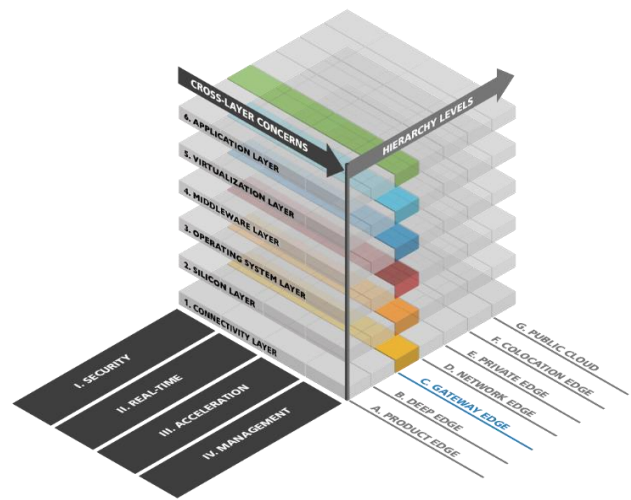

**Figure 1**. Reference Architecture Model for Edge Computing (RAMEC) developed by the Edge Computing Consortium Europe, retrieved from: https://ecconsortium.eu/

We see Edge Computing as a distributed cloud computing paradigm that is potentially applicable for all areas of technological use cases. In order to provide orientation and to put activities such as initiatives, standards or implementations into a broader context, a reference model is needed. For this purpose, within the Smart Grid context the Smart Grid Architecture Model (SGAM) and within the Smart Manufacturing context the Reference Architecture Model Industry 4.0 (RAMI 4.0) have been developed. Best to our knowledge, the preliminary Reference Architecture Model for Edge Computing (RAMEC, Figure 1) is the detailed model that spans a suitable overview of the multi-dimensional Edge Computing problem space. Therefore, this study uses the RAMEC model as a framework for the analytical part and for the sustainability roadmap. It has been developed by the industry-oriented Edge Computing Consortium Europe (ECCE), which is currently in the process of being founded. The RAMEC shows how Edge Computing can be technically distinguished according to concerns and hierarchy levels, i.e. at which kind of "edge" the computing capability is located.

The model suggests that the "edge of the network" is not a clearly defined localization of data processing but rather depends on the specific use case. Additionally, the technical layers enabling Edge Computing capabilities can be distinguished. In the interplay of these layers and hierarchy levels, several cross-concerns emerge. In these areas many current challenges and open issues on Edge Computing can be located.

## 3  Research Design and Method

To approach this young field of edge computing development, we carried out an open and mixed methods study including qualitative interviews [24] and explorative content analysis. An explorative content analysis [25] is characterized by a growing list of the values for each variable summarized into larger categories. In this way, the manual access and a review process of the online textual materials allows to carry out a comprehensive survey including all apparent characteristics of the study objects which might be missed out in an automated analysis.

The initial sample of Edge Computing initiatives was based on interviews of two experts in the field of Edge Computing from the ECCE. To extract the largest sample as possible, we decide on a broad operationalization of Edge Computing and include related developments around the Internet of Things and Fog Computing. The data sample has been collected by keyword-led online search using snowball sampling (i.e., following Web links to related resources) and the keywords: "[edge OR fog OR distributed] computing", "iot", "internet of things", "iiot", "industry ["4.0" OR 40]", "multi-access edge computing", "mobile edge computing". Afterwards the extracted initiatives have been cleaned according to our Edge Computing definition. A total sample of 75 initiatives has been collected to answer the research question "How can current Edge Computing initiatives be characterized?"

An open online search and snowball sampling on the 75 collected initiatives resulted in the collection of 88 primary resource materials: 56 descriptive sites (e.g. "about" pages,), 9 PR publications (e.g. white paper, studies), 9 news/press releases, 7 branch media articles (from media like Computer Base), 7 community resources (e.g. Wikipedia articles, online documentation platforms). We analyzed the materials according to the following qualitative variables (see [25] for methodological details):

- Working area: On which technical aspect an initiative aims to contribute to.
- Branch: In which branch or sector the initiative operates
- Type: What is the type of an initiative
- Harmonization: A norm or standard that has been recognized by a standardization body or norming institution, and that possesses an identification code
- Software project: A community-based software development project, a new company department or research institute that is dedicated Edge Computing
- Consortium: an association of two or more companies, organizations or governments
- National Initiative: A political-economy driven incentive, like a campaign or a financial boost to support the economic development into a certain direction
- Sustainable development: which perspective(s) of Sustainability can be found
- Location: where or in which country an initiative is located

- Hierarchy level: to which hierarchy level an initiative can be allocated
- Layer: to which technical development layer an initiative can be allocated

## 4      Results

To answer the research question "How can current Edge Computing initiatives be characterized?", we examined outlets of the initiatives themselves as well as branch media articles, and community resources. By this qualitative research which some quantitative elements, we receive direct insights on the field, which serve to describe the sample and hence the research object "Edge Computing initiatives".

The working areas of the surveyed Edge Computing initiatives have a wide range (see Table 1). The most occurring ones are: Edge computing (unspecified), mobile edge computing, orchestration, and platform development.

**Table 1.** Working areas of the initiatives. Selection of working areas with frequency (F) > 1

| *Working Area* | *F* | *Working Area* | *F* | *Working Area* | *F* |
|---|---|---|---|---|---|
| Edge computing (unspecified) | 8 | Network function virtualization | 3 | Architecture | 2 |
| Mobile edge computing | 6 | Virtualization platform | 2 | Applications | 2 |
| Orchestration | 5 | Cloud edge computing | 2 | Container | 2 |
| Platform development | 5 | Distributed cloud | 2 | Software | 2 |
| Cloud computing | 3 | Virtualization | 2 | OPC UA | 2 |
| Fog computing | 3 | Microservices | 2 | PLC | 2 |
| Management | 3 | Requirements | 2 | TSN | 2 |
| Framework | 3 | Infrastructure | 2 | | |

**Table 2.** Branch/domains of the current EC initiatives. Reduced sample of n = 39 since most national initiatives and software projects cannot be allocated to a specific branch or domain.

| *Branch/Domain* | *F* | *Branch/Domain* | *F* | *Branch/Domain* | *F* |
|---|---|---|---|---|---|
| Telecommunication | 11 | Enterprise | 3 | Consumer | 2 |
| Industrial | 9 | Transportation | 2 | Energy | 2 |
| Manufacturing | 5 | Automotive | 2 | Meteorology | 1 |

The analysis confirms that the telecommunication sector as well as the industrial and manufacturing sectors are most prevalent in the Edge Computing development (see Table 2). However, the largest part of initiatives (36 of 75) cannot be allocated to a certain branch or domain. Software projects are mainly tackling basic technical development work that is not yet connected to a particular field of application. The national initiatives generally aim at strengthening the economy of a country by promoting technological developments such as Edge Computing, they do not limit this promotion to a certain sector.

The distribution of the types of Edge Computing initiatives has been found as follows: 22 harmonization organizations, 28 software projects, 20 consortia, and 16 national strategies (see Figure 2). In total, 22 of the 75 initiatives presents sustainability concerns in their resource materials. A large part of the surveyed initiatives is internationally organized (36), one among them is a European initiative. The others have an allocation to a certain country: Germany (13), US (7), France (3), UK (2), Japan (2), China (1), Spain (1), Korea (1), Taiwan (1), Switzerland (1), Poland (1), Italy (1), New Zealand (1), Austria (1), and the Netherlands (1).

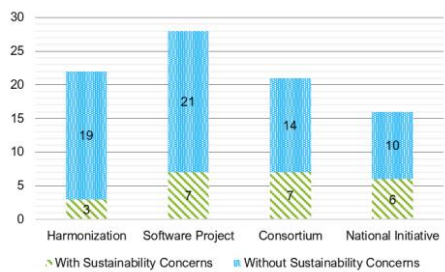

**Figure 2**. Edge Computing initiative types and addressing of sustainability concerns, n = 75

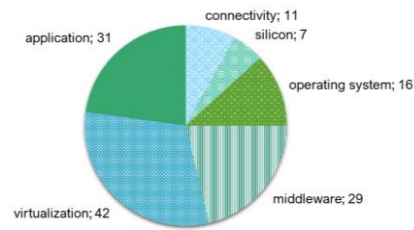

**Figure 3.** RAMEC layers addressed by the current Edge Computing initiatives, n = 55; multiple options allowed

The sampled initiatives could not easily be allocated into the hierarchy levels and layers of the RAMEC model. In many cases, the Edge Computing initiatives do not describe which edge level they are focusing on, so the variable "hierarchy level" could not be answered. Nevertheless, the categorization according to the RAMEC layers was possible and showed the prevailing layers of virtualization, application, and middleware (see Figure 3). Based on these empirical insights we move on to tackle the second ambition of the study: to provide a roadmap for sustainable Edge Computing.

## 5  Discussion

### 5.1  Comprehensive Survey of Current Edge Computing Initiatives

The research question, how current Edge Computing initiatives can be characterized can be answered as follows: In this explorative study 75 initiatives have been collected; most initiatives are software-focused projects of which the majority is internationally organized (17 of 27). The 75 initiatives include 28 software projects, 22 harmonization projects, 20 consortia, and 16 national initiatives. The leading working areas are mobile edge computing, orchestration, and platform development.

We found that a small portion of 22 of the 75 initiatives of all types (see Figure 2) show intentions for a sustainable development of Edge Computing. This number can generally be regarded as low and needs to be extended in future. Even for the initiatives that express sustainability aims, we criticize that this is mainly done in a rather vague and reticent way, e.g. "[…] new and better driving experiences with safer passage, smoother traffic flows and optimized energy consumption for lower emissions that are kinder to the environment." Here, we suggest addressing all dimensions of sustainability and to be more precise in how these intentions should be achieved. To unfold these numerous steps to a Sustainable Edge Computing, we suggest the following Roadmap for Sustainable Edge Computing.

The results indicate that the Edge Computing paradigm might broadly be shaped by telecommunication, industrial and manufacturing communities, which might cause adoption issues for other branches and sectors, for example healthcare or traffic systems, leveraging Edge Computing for their purposes. The search terms used in our survey, e.g. "Industry 4.0" or "mobile edge computing", might have led to an overrepresentation of the industrial and telecommunication sectors in the depiction of the Edge Computing development, however it equally represents the status quo of the economic development. The ignorance of such search terms would might have led to an unrealistic sample. Nevertheless we acknowledge the limitation of the study that due to the data collection method by keyword-led online research and snowball sampling, the analyzed sample is not exhaustive and therefore of limited representativity. Moreover, the sample is probably biased towards the German context as one of the search terms "Industry 4.0" originates from a German policy, but we find also that this term is increasingly used in an international context.

Due to a lack of details in the examined resources, it was not possible to position the Edge Computing initiatives in the RAMEC model. We see this fact as an issue that needs to be addressed and therefore stress the importance of a common reference architecture model for Edge Computing. We found indications by the initiatives just for one of the RAMEC dimension: the development layers. The majority of initiatives is dealing with the virtualization layer whereas; the silicon and connectivity layers are the least addressed and could represent a gap of development.

### 5.2 Roadmap for a Sustainable Edge Computing Development

The sustainability concept was defined by the United Nations as human development that would meet present needs without compromising the ability of future generations to meet their own needs [19]. Following the Brundtland report of the World Commission on Environment and Development of the United Nations, the Enquete Commission of the German Bundestag "Protection of Man and the Environment" describes sustainability as the concept of a sustainable development of the economic, ecological and social dimension of human existence [20]. We use these three dimensions in our model to reveal the crucial stages where Edge Computing can support or corrupt sustainable development. We use the Edge Computing concerns from RAMEC and relate them to the sustainability dimensions.

Sharing the claims for a need of sustainable computing, we propose a sustainability roadmap for Edge Computing (see Figure 4). The roadmap reveals many traps of an exploitative and non-sustainable development, but it shows at the same time several opportunities how Edge Computing can contribute to a more sustainable digital development. In the following, all boxes of the roadmap will be explained in detail.

**Security and privacy issues.** Edge Computing encompasses multiple security and safety issues. It needs to be resilient and robust against criminal attacks and of accidental uses. Edge Computing applications that work in a safe and secure way can ensure fair processes in which all people are handled in the same way (Box 1A), e.g. surveillance cameras that analyze facial images on the spot, which can improve social stability. The fact that data can be analyzed and stored close to the data sources solves several privacy-related issues as the data must not be transported anymore to a central server. People's data are therefore protected as they cannot be centrally stored and combined anymore which is beneficial for a socially sustainable development being data protection- and privacy-compliant. However, permanent surveillance is a privacy-related issue especially when the processes behind it are not transparent for a society or even undemocratic.

| Dimensions of Sustainability \ Edge Computing Concerns | A — Social Sustainability | B — Ecological Sustainability | C — Economic Sustainability |
|---|---|---|---|
| 1. Security & Privacy Issues | (+) Fairness and social stability  (-) Self-determination and personal freedoms | (-) Increase of energy consumption through security measures | (+) New business models supported by norms and regulations  (-) Cybercrime |
| 2. Real-Time Capabilities | (+) Responsiveness  (-) Ubiquity | (+) Reduction of energy consumption through process optimization  (-) Increase of energy consumption through data processing | (+) Competitive advantage  (-) Stable control loops |
| 3. Learning and Smart Capabilities | (+) Flexibility and autonomy  (-) Bias on social groups, discrimination or exclusion of individuals | (+) Reduction of energy consumption through process acceleration  (-) Increase of energy consumption through training | (+) Economy of scale  (-) Data validity and quality |
| 4. Management Capabilities | (+) Monitoring and traceability  (+) Variability  (-) Accessibility | (+) Monitoring and traceability  (+) Use of renewable energies  (-) Material consumption and recycling of devices and sensors | (+) Monitoring and traceability  (+) Standards and interoperability |

(+) Rather beneficial
(-) Rather detrimental

**Figure 4**. Roadmap for Sustainable Edge Computing development

Especially when deployed in environments such as traffic or industrial production, edge computing requires appropriate encryption and authentication depending on the criticality (Box 1B). In some cases, it requires a relatively small device or even single sensor to tunnel for example over TLS [26]. Considering the massive amount of IoT devices and sensors due to their ubiquity, this would require increased energy consumptions compared to unsecured devices. Such increase preferably needs to be addressed by renewable energies, e.g. by energy harvesting [27]. Related to this, as renewable energy will be sourced in a distributed manner and the distribution of energy from multiple small sources is challenging, the discussed distributed Cloud Computing paradigm (Edge Computing) can inherently be used to consume energy where it is generated and therefore reduce energy supply problems at centralized data centers. Further, the choice of encryption can save resources. Suárez-Albela et al. [28] examined that securing IoT scenarios by Elliptic Curve Cryptography (ECC) can save up to 50 % of the energy consumption compared to securing it by Rivest-Shamir-Adleman (RSA).

As the computing capability is carried out in a distributed network, it is more difficult to ensure the same security standards in a decentralized computing architecture than in a centralized computing center [29] (Box 1C). The consistent and forceful implementation of respective standards would improve the security status of many actors in the field and hence the conditions for their customers. Security innovations for Edge Computing could moreover be the basis for new business models and sustainable digital economy. Diversity in component, service and application supplies would not only be advantageous for customers and business partners, but also improve security since very same errors will not be multiplied. The lack or limited technical depths of security norms could however also be seen as an invitation for cybercriminals, which needs to be addressed as soon as possible.

**Real-time capabilities.** The real-time potentials and low latency properties of Edge Computing are the main drivers for its development. The increased responsiveness of edge applications compared to cloud applications represents a social advantage as user requests can be addressed in real-time. However, the ubiquity of IoT and Edge Computing could lead to permanent monitoring and scoring which might be detrimental for people's freedoms (Box 2A).

With regard to real-time applications it must be planned holistically if a real-time data transfer enables the reduction of energy consumption or if it leads in the end to a higher energy consumption (Box 2B). For example, if smart streetlights indicate in real-time free parking slots in a city and contribute to reduce the search for parking slots in traffic-jammed and smoggy cities, it would be beneficial for sustainable development. However, if such an application would be rolled out in any traffic light by default, also in low frequented areas, permanent data analysis and storage would cause unnecessary resource and energy consumption.

Some innovations, especially in the industrial or mobility sector, require an extremely fast and reliable way of data processing which physically cannot be solved anymore in the cloud or via mobile data (Box 2C). Offering Edge Computing solutions would hence be a great competitive advantage to previous cloud computing solutions. For example the computing capabilities on smart phones can be used to carry out local vision-based human activity recognition [30]. However, Edge-Computing-reliant control circuits steering processes like micro-trading in the financial sector can become unstable if reactions happen too fast or have to be readjusted. As control circuits require time for leveling their processes, the implementation of real-time solutions could trouble the stability of such control circuits.

**Learning and smart capabilities.** Edge devices allow possibilities to enhance data collection and analysis. Bringing semantics into the data makes it easier to recognize similar data from multiple sources, for example, very small edge devices can simply suppress redundant data, while medium or larger devices provide application-specific software. Some edge technologies allow code mobility or even the ability to deploy on edge devices the same containers deployed in the cloud [26]. From a social perspective, caution is advised: The use of people related training data for predictions includes a significant risk of biases for social groups and the respective discrimination and exclusion (Box 3A). Several examples have reached media attention where machine-made algorithm-based decisions have significantly discriminated people, and hence amplify racist and sexist behaviors [31].

Learning applications can be leveraged for ecologic sustainability (Box 3B), for example to optimize and reduce energy consumption in households and buildings [32]. From a holistic perspective, one must not only compare the energy consumption before and after the implementation, but also the energy consumption of creating and training of a learning algorithm, as well as the storage of the training data. A benefit for sustainable development would appear if the scale of the reduced energy consumption enabled by one trained algorithm is larger than the energy this algorithm has consumed to be created and trained.

The implementation of learning and smart capabilities might enable effects of economics of scale (Box 3C.). Once an algorithm reliably predicts certain behaviors this mechanism might be used for other use cases, and hence the efficiency of numerous productions or services might be increased without taking more investments. For example, a smart traffic light system that has been expensively developed by one city to improve the urban traffic flow could be sold as a product to other cities without significant changes. Yet, one must be aware of the risks appearing if the training data is not valid or reliable. For example, an automated access system using facial recognition software might work perfectly in a developer training environment, but it might not work in a real-world scenario where wheelchair users or children might not be able to use an interface created for adults standing in front of a door.

**Management capabilities**. The promotion and successful implementation of the Edge Computing paradigm is a large management task. Edge-enhanced management procedures could be problematic in terms of accessibility (Box 4A). Automatically managed systems without close human support could pose problems for people with special needs, e.g. blind people could not follow written instructions to operate an edge application; children or wheelchair users could be unable to access an interface.

Not only the computing capability should be decentralized but also the power supply (Box 4B.). Decentralized power supplies might leverage renewable energies on the spot. For example, applications in urban areas could be powered by solar panels or wind turbines. Smart factories could be connected to a water power plant close to their location. At the same time, it might be considered that the digitalization of formerly analogue processes will potentially consume more energy. Scholars moreover emphasize the consideration of material consumption particularly plastics and rare metals [33, 34] as well as the recycling problems and global health issues caused by e-waste [35].

Taking into account the RAMEC model cited in Section 2.4, the management of an Edge Computing solution must be considered at any layer and hierarchy (Box 4C). This concerns standards and interoperability mainly related to Edge Computing initiatives' agreements but also to applications in general. This task does not only encompass the technological solutions but also collaboration of companies and the implications on business-making.

**Monitoring and Traceability.** A crosscutting surplus of Edge Computing for social, ecological and economical sustainability are the potential improvements in measuring and monitoring indicators for the sustainability dimensions, which can be used to trace back and forth between root causes and quantitative and qualitative impacts. Monitoring can be the basis for adaptive sustainability governance in order to be able to anticipate and prepare for non-linear, erratic or unexpected changes in increasingly complex societal contexts [36].

A thorough consideration of these sustainability concerns complicates the estimation if an Edge Computing application is worth being developed in view of sustainability. However, it is necessary to carry out such systemic considerations in order to ensure that Edge Computing does not only become one more accelerant of economic growth but contributes to social and ecological well-fare. This roadmap is a first proposal for Sustainable Edge Computing and will be further elaborated. It is neither exhaustive nor complete and seeks for revisions and further discussion.

## 6      Conclusion

In this explorative study, 75 Edge Computing initiatives have been analyzed. Most initiatives are software-focus projects of which the majority is internationally organized (17 of 27). The 75 initiatives include 28 software projects, 22 harmonization projects, 20 consortia, and 16 national initiatives. The leading working areas are mobile edge computing, orchestration, and platform development.

The results indicate that the Edge Computing paradigm might be broadly shaped by telecommunication, industrial and manufacturing communities which might cause implementation issues for other branches and sectors, for example healthcare or traffic systems, leveraging Edge Computing for their purposes.

Further, we adhere the argument that nowadays research and development should include sustainability concerns in their work routine in order not to further exceed the natural boundaries of our planet. One important finding of our empirical analysis is that sustainable developments generally receive too little attention within the framework of Edge Computing. We deduce that this shortcoming might be caused by the strong linkage of Edge Computing to the "old-fashioned sectors" of industrial and manufacturing contexts. We therefore developed a roadmap for Sustainable Edge Computing relating the current technical issues with the three sustainability dimensions. We hope that the developer's and managers around Edge Computing are building their future achievements on such a roadmap and therefore propose our model (see Fig. 4) as a first step towards Sustainable Edge Computing.